\documentclass[pre,twocolumn,a4paper,showpacs]{revtex4-1}
\usepackage{amssymb}
\usepackage{wasysym}
\usepackage{graphicx}
\usepackage{epsfig}
\usepackage{subfigure}
\begin{document}

\title{Effect of long-range interactions on the phase transition of Axelrod's model}

\author{Sandro M. Reia and Jos\'e F. Fontanari}
\affiliation{Instituto de F\'{\i}sica de S\~ao Carlos,
  Universidade de S\~ao Paulo,
  Caixa Postal 369, 13560-970 S\~ao Carlos, S\~ao Paulo, Brazil}


\begin{abstract}
Axelrod's model with  $F=2$ cultural features, where each feature can  assume $k$  states drawn from a Poisson distribution of parameter $q$,  exhibits a  continuous   nonequilibrium phase transition in the square lattice. Here we use extensive Monte Carlo simulations and finite size scaling to study the critical behavior of   the order parameter $\rho$, which  is the fraction of sites that belong to the largest domain of an absorbing configuration averaged over many runs. We find that it vanishes as $\rho \sim \left (q_c^0 - q \right)^\beta$  with $\beta  \approx  0.25$ at the critical point $q_c^0 \approx  3.10$  and that the exponent that measures the width of the critical region is  $\nu^0  \approx 2.1$. In addition, we find that introduction of  long-range links by rewiring the nearest-neighbors links of the square lattice with probability $p$  turns the transition discontinuous, with the critical point $q_c^p$ increasing from $3.1$ to $27.17$, approximately,  as $p$ increases from $0$ to $1$. The sharpness of the threshold, as measured by the exponent $\nu^p \approx 1$ for $p>0$,  increases with the square root of the number of nodes of the resulting
small-world network. 


\end{abstract}

\maketitle

\section{Introduction}\label{sec:Intro}

Axelrod's model for the dissemination of culture takes into account two key ingredients of social dynamics \cite{Lazarsfeld_48,Castellano_09,Galam_12}, namely,  
social influence through which people become more similar when they interact  and homophily, which is 
the tendency of individuals to interact preferentially with similar others \cite{Axelrod_97}.  Explicitly, in Axelrod's   model the individuals are modeled by   agents 
which are   strings  of cultural features of length $F$, where each feature can adopt a certain number $k$ of distinct states. The $F$ cultural features of an agent determine its culture. 
The $N$ agents are fixed at the nodes of a network (regular lattice or complex network) and the interaction between  two connected agents takes place
with probability proportional  to the number of states they have in common  and  always results in an increase of the similarity
between them.  

Whereas social influence is a main  feature of the standard two-opinion voter model \cite{Ligget_85} and homophily has been allowed for  in the three-opinion constrained voter model \cite{Vazquez_03,Vazquez_04}, the number of states -- two and three, respectively -- is not a free parameter in those  models. As a result, the two-opinion voter model exhibits only consensus absorbing configurations   in regular lattices of  arbitrary dimension \cite{Ligget_85} (see, however, \cite{Castellano_03} for a report of incomplete ordering on small-world networks). The same conclusion holds for the constrained voter model in  an infinite  one-dimensional lattice  (see, e.g., \cite{Lanchier_12a,Lanchier_12b,Vilone_02,Biral_15}) and  since increasing the range of the  agents' interactions favors the consensus  regime \cite{Greig_02,Klemm_03a}, we expect this conclusion to hold for regular lattices
of higher dimension  as well.
In Axelrod's model, however,   increase of  the number of states $k$   for a fixed string length $F$ leads the social dynamics  to freeze in multicultural absorbing configurations even in a one-dimensional lattice \cite{Lanchier_12b,Vilone_02}.

In fact,   Axelrod's model exhibits two types of absorbing configurations in the thermodynamic limit $N \to \infty$: ordered  configurations, which  are characterized by a few cultural domains of macroscopic size $\mathcal{S}$, and  disordered  configurations, where all domains are microscopic
\cite{Castellano_00,Klemm_03,JEDC_05,Vazquez_07,Peres_15}. By cultural domain we mean  a bounded region of uniform culture. 
The competition between the disorder of the initial configuration that favors cultural fragmentation and  the ordering bias of social influence 
that favors homogenization results in the nonequilibrium phase transition between those two classes  of absorbing states  \cite{Castellano_00}.  We note that, similarly to the   standard percolation \cite{Stauffer_92},  the phase  transition occurs in the properties of the absorbing states and so it  is static in nature \cite{Vilone_02}.

In this paper we reexamine the nonequilibrium phase transition of a variant of Axelrod's model in which the initial states of the $F$  cultural  features of the  agents are  drawn randomly  from a Poisson distribution of parameter $q \in \left [ 0, \infty \right )$,
\begin{equation}\label{Poisson}
P_k = \exp \left ( -q \right ) \frac{q^k}{k!}
\end{equation}
with $k=0,1,2, \ldots$. In the original model, these states are chosen randomly from a uniform distribution on the integers $1, 2, \ldots, q$
\cite{Axelrod_97}.
In the square lattice, the Poisson variant exhibits a continuous phase transition for $F=2$   and a discontinuous one for $F > 2$ \cite{Castellano_00}.   In the one-dimensional lattice, only the disordered regime exists for  $F=2$ and a  discontinuous transition between the disordered and the ordered regimes is observed for $F>2$ \cite{Vilone_02}.  Here we  focus on the case $F=2$ only and study the effect of long range interactions on the standard order parameter of Axelrod's model $\rho = \left \langle \mathcal{S}_{max} \right \rangle/N$  that measures the  fraction of agents that belong to  the largest cultural domain, whose size is denoted by $\mathcal{S}_{max}$, averaged over many independent runs   (see, e.g, \cite{Castellano_00,Vilone_02,Klemm_03a}).

A previous study of the  continuous phase transition for $F=2$ in the square lattice considered as order parameter  the mean density of domains   $\mu = \left \langle \mathcal{N} \right \rangle/N$, where $\mathcal{N}$ denotes the number of domains of an absorbing configuration,  and showed that it vanishes as $\mu \sim \left ( q - q_c^0 \right)^{\beta'}$  with $\beta' = 0.67 \pm 0.01$ at the critical point $q_c^0 = 3.10 \pm 0.02$ \cite{Peres_15}. We note that $\mu$ is not an order parameter for the standard percolation since it is continuous and non-zero  at the threshold  \cite{Stauffer_92}.
 The advantage of considering $\mu$ is that it is nonzero in the disordered regime  where  the convergence of the dynamics to the absorbing configurations is very fast as compared to the convergence to the quasi-consensus configurations of the ordered regime
 (see, e.g., \cite{Biral_15}).  The present study  complements that analysis by showing that the more usual order parameter $\rho$ vanishes at the critical point  as $\rho \sim \left ( q_c^0 - q  \right)^{\beta}$  with $\beta = 0.25 \pm 0.02$. 
 
 In addition, here  we study the effect of long range interactions  by considering small-world networks in which each link of the square lattice with periodic boundary conditions
  is rewired with probability $p$. Rewiring of a link is done by replacing the original neighbor of a given site by a random site chosen uniformly among  all possible sites that avoid self-loops  and link duplication \cite{Watts_98}. The average degree of the resulting network is
 4 regardless of the value of $p$. The extremes  $p=0$ and $p=1$ correspond to the regular square lattice  and to  a random network, respectively.  Since the underlying regular lattice used to generate the small-world networks is the square lattice we can introduce the length $L \equiv N^{1/2}$, although it has  no geometrical meaning for $p > 0$.   We find that the transition is discontinuous for $p> 0$  and that  the  critical exponent that determines the width of the critical region for finite $L$ is $\nu^p = 1.0 \pm 0.05$. However, in the case $p=0$ where  the  transition is continuous  we find $\nu^0 = 2.1 \pm 0.1$.

 The remainder of the paper is organized as follows. For the sake of completeness,  in Section \ref{sec:model} we present a brief account of Axelrod's model \cite{Axelrod_97} and of the
 Watts and Strogatz algorithm for constructing small-world networks \cite{Watts_98}.  In Section \ref{sec:op} we study the behavior of the order parameter $\rho$ near the critical region for three network topologies: the square lattice ($p=0$), random networks ($p=1$) and small-world networks with rewiring probability $p=0.1$.
 Finally, Section \ref{sec:conc} offers our concluding remarks.

 \section{The Model}\label{sec:model}
 
The Poisson variant of Axelrod's  model differs from the original model only by the procedure that generates the cultural 
states of the agents at the beginning of the simulation: for each feature $l=1, \ldots, F$ of each agent $i=1, \ldots, N$ a state  $k=0,1, \ldots$ is
drawn  independently  using the distribution (\ref{Poisson}). Once the initial configuration is set, the dynamics  proceeds
as in the original model \cite{Axelrod_97}. In particular,   at each time we pick an agent at random
-- the target agent -- as well as one of its neighbors. These two 
agents interact with probability equal to  their cultural similarity, defined as the fraction of 
common cultural features they have.  An interaction consists of selecting at random one of the distinct features, and making the
selected feature of the target agent equal  to  the corresponding  feature of its neighbor. This procedure is repeated until 
the system is frozen into an absorbing configuration.  According to these rules, at an absorbing configuration  any pair of neighbors are either identical  or completely different  regarding their cultural states.

A feature that sets Axelrod's model  apart from most
lattice models that exhibit nonequilibrium phase transitions \cite{Marro_99} is  that all stationary 
states of the dynamics are absorbing states, i.e., the dynamics always freezes in one of these states. This
contrasts with lattice models that exhibit an active state in addition to infinitely many  absorbing states \cite{Jensen_93}
and the phase transition occurs between the active state and the usually equivalent absorbing states (see \cite{Pace_14} for
a simple change in the update rule of Axelrod's model that results in dynamically active metastable states).

The implementation of  the Watts and Strogatz algorithm \cite{Watts_98} for constructing the small-world networks used in our study  begins  with  a  square lattice of linear size $L$ with nearest neighbors interactions and periodic boundary conditions (i.e., a torus).  Then for every site  $i= 1,\dots, N=L^2 $  we rewire the link between $i$ and, say, its left neighbor, with probability $p$. As mentioned before, rewiring of a link is done by replacing the original neighbor of site  $i$ (in this case, the left neighbor)  by a random site chosen uniformly among  all possible sites that avoid self-loops  and link duplication. The procedure is then repeated  for, say, the top neighbor of every  site of the lattice.
Note that for $p=1$ the resulting network is not a classic random network  \cite{Gilbert_59}, since the rewiring scheme guarantees that any site will be connected to at least  $2$ other sites,  whereas for  classical random graphs  any site  has  probability $e^{-4}$ of being  isolated from the
other $N-1$ sites. Otherwise, the resulting networks are very similar to classical  random graphs.

 \section{Analysis of the order parameter}\label{sec:op}

For a randomly generated initial configuration of the  agents' cultures, we follow the dynamics of Axelrod's model until it reaches an absorbing state -- this comprises a single run -- and then we calculate the size of the largest  cultural domain
$\mathcal{S}_{max}$.  Average of this quantity over a large number of independent runs (typically $10^4$), which differ by the choice of the initial cultural states of the agents as well as by their  update sequence, yields the order parameter $\rho = \left \langle \mathcal{S}_{max} \right \rangle/N$ we use to characterize the nature of  the absorbing configurations. In the case of random ($p=1$) and small world ($0<p<1$) networks we generate a different network  for each run.

\begin{figure}[!ht]
  \begin{center}
\includegraphics[width=0.48\textwidth]{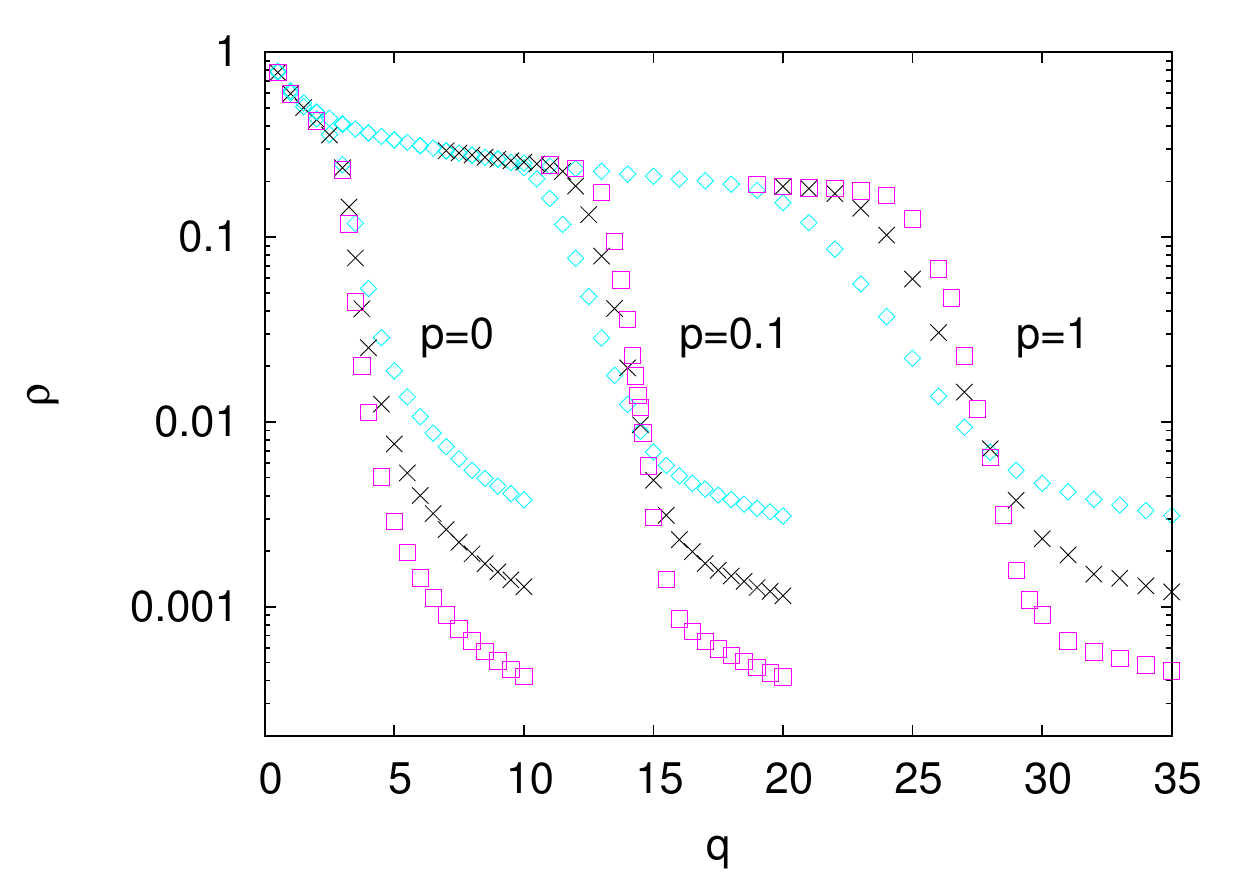}
  \end{center}
\caption{(Color online) Mean fraction of sites in the largest domain $\rho$  as function of the Poisson parameter $q$  for the square lattice ($p=0$),
small-world networks with $p=0.1$ and random networks ($p=1$). The different symbols represent different number of sites $N=L^2$, viz.  $L = 100$  ($\Diamond$), $L =  200 $ ($\times$) and  $L=400$ ($\square$).   The error bars are smaller than the symbol sizes.
 }
\label{fig:1}
\end{figure}

Figure \ref{fig:1} offers a bird's eye view of the dependence of the order parameter $\rho$ on the parameter $q$  of the Poisson distribution, which determines the number of different cultural states in the initial population, for the three network topologies we will consider here, viz.
the square lattice ($p=0$), small-world networks with $p=0.1$ and random networks ($p=1$). This figure reveals a few interesting features of  
Axelrod's model. First, the presence of long range links favors the ordered regime as indicated by the shift of the threshold region to high values of $q$, i.e., as the fraction $p$ of long-range links increases, the disorder in the initial configuration must also increase in order the dynamics  reaches a disordered absorbing configuration \cite{Klemm_03a}. Second, deep into the ordered phase, say for $q <10$, 
the order parameter is completely insensitive to increasing  the fraction of long-range links beyond a certain value, say $p=0.1$. Third, the crossing of the data for different $L$ for $p>0$ signals the presence of a discontinuous transition in the limit $L \to \infty$, whereas  for $p=0$ the  condition
$\rho \left ( L_1 \right )  \geq \rho  \left ( L_2 \right ) $ for $L_1 < L_2$ is satisfied for all values of $q>0$.

Next we study in detail the behavior of the order parameter $\rho$ in the  critical region for the three network topologies exhibited in Fig.\ 
 \ref{fig:1}.
 
 \subsection{Square lattice}
 
 The square lattice considered here exhibits only short range interactions and so it offers a baseline  for  assessing the effects of long-range links on the phase transition of Axelrod's model. In addition, as pointed out before, this study complements  previous analyses of the Poisson variant of Axelrod's  model in the square lattice \cite{Castellano_00,Peres_15}. In fact, whereas Ref.\  \cite{Castellano_00} considered the qualitative aspects of the continuous nonequilibrium transition (in the sense  that  there were  no attempt to  estimate  the critical point $q_c^0$ and the critical exponents),  Ref.\  \cite{Peres_15} focused on an alternative order parameter. Since we expect that the location of the critical point  is not affected by the choice of the order parameter (in cases there is such choice),  here we borrow from  \cite{Peres_15} the estimate $q_c^0 = 3.10 \pm 0.02$. The discernment of this resolution will be evaluated by the goodness and consistency of our results. 
 
\begin{figure}[!ht]
  \begin{center}
\includegraphics[width=0.48\textwidth]{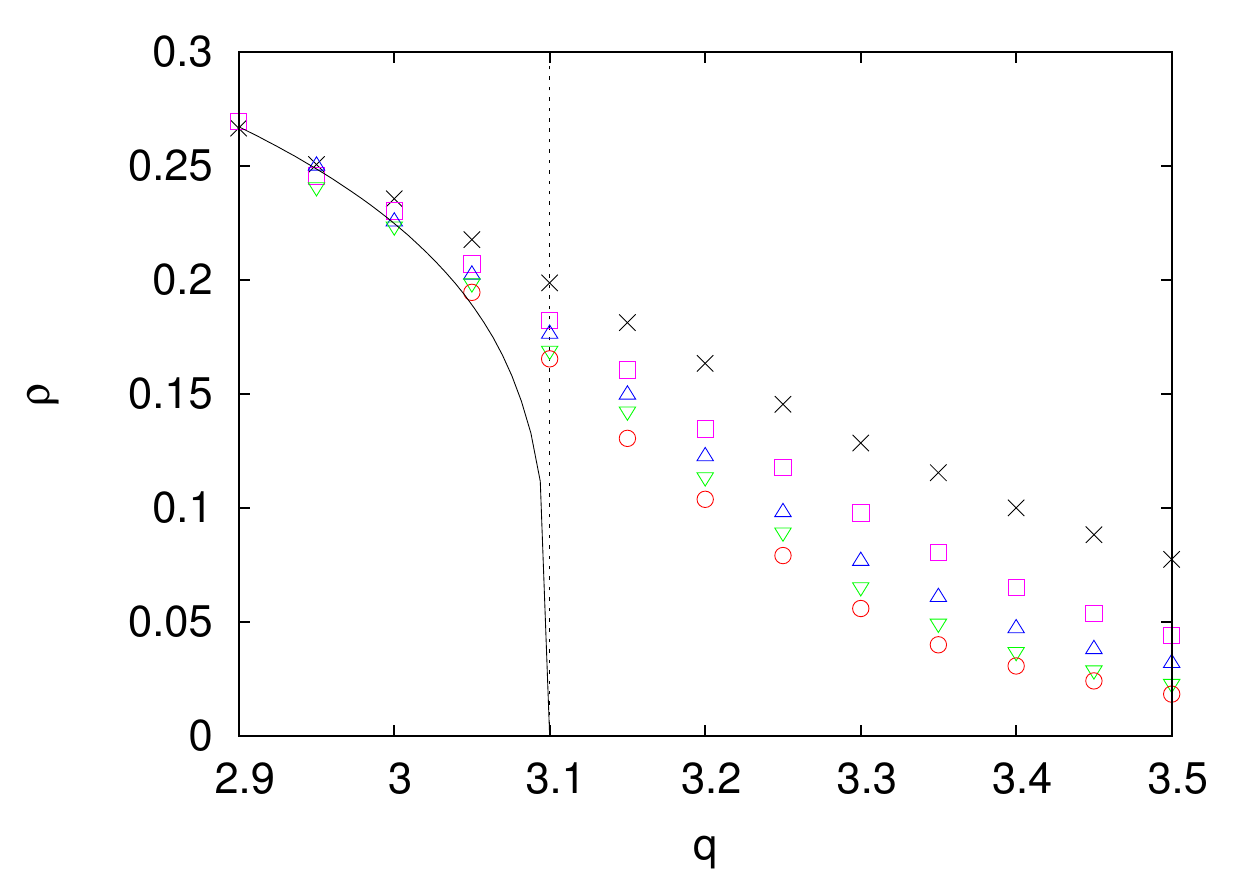}
  \end{center}
\caption{(Color online) Mean fraction of sites in the largest domain $\rho$  in the critical region  for the square lattice ($p=0$) with
linear  size $L =  200 $ ($\times$), $L=400$ ($\square$), $L=600$ ($\triangle$), $L=800$  ($\triangledown$), and $L=1000$ ($\bigcirc$).   
The dotted vertical line indicates the location of $q_c^0 \approx 3.1$ and the solid curve is the two-parameters fitting  function  $\rho= \mathcal{A} \left ({q}_c^0 - q \right )^{\beta}$ with $\mathcal{A}=0.40 \pm 0.01$,  and $\beta= 0.25 \pm 0.02$.
The error bars are smaller than the symbol sizes. }
\label{fig:2}
\end{figure}

Figure \ref{fig:2} offers a more detailed view of the order parameter $\rho$ near the critical point $q_c^0$. Rather remarkably,
the finite size scaling theory asserts  that   for  large $L$  the data shown in this figure can be described by the scaling relation \cite{Privman_90}
\begin{equation}\label{scal1}
\rho \sim L^{-\beta/{\nu^0}} f_0 \left [ L^{1/{\nu^0}} \left ( q_c^0 -  q \right ) \right ],
\end{equation}
where the scaling function is $f_0 \left ( x \right ) \propto x^\beta$ for $x \gg 1$ and $\nu^0 > 0$ is a critical exponent that determines the width of the critical region for finite $L$.  Hence in the limit $L \to \infty$ one has $\rho \sim  \left (  {q}_c^0 - q \right )^{\beta}$ near the critical point, where $\beta > 0$ is a critical exponent.  

\begin{figure}[!ht]
  \begin{center}
\includegraphics[width=0.48\textwidth]{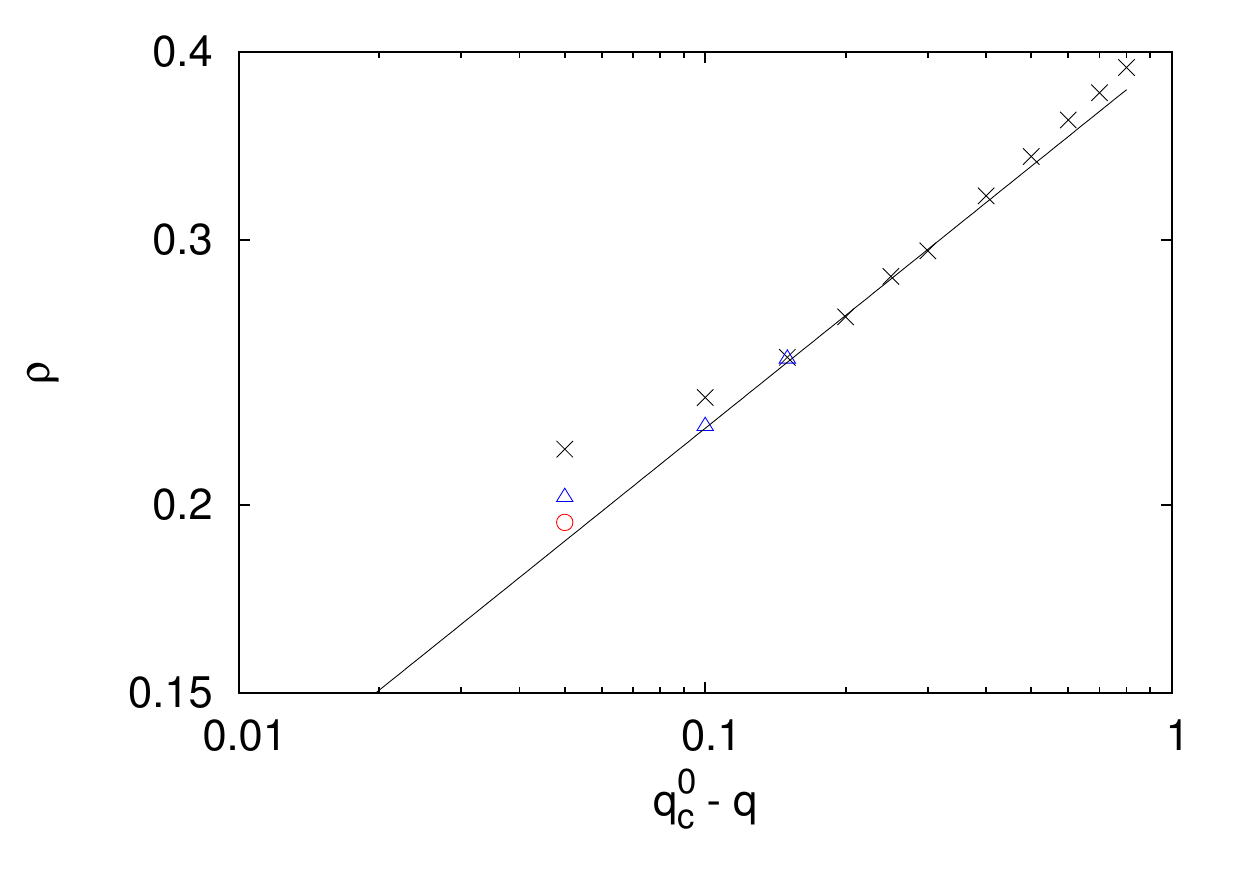}
  \end{center}
\caption{(Color online) Log-log plot of $\rho$ against $q_c^0-q$ with $q_c^0= 3.1$ for the square lattice ($p=0$) with  $L =  200 $ ($\times$), $L=600$ ($\triangle$) and $L=1000$ ($\bigcirc$).  The solid line is the fitting  function $\rho= \mathcal{A} \left ({q}_c^0 - q \right )^{\beta}$ with $\mathcal{A}=0.40 \pm 0.01$,  and $\beta= 0.25 \pm 0.02$. }
\label{fig:3}
\end{figure}

Figure \ref{fig:3} summarizes the results of the  fitting of the data for $L=200$ with the function $\rho = \mathcal{A} \left ( {q}_c^0 - q \right )^\beta$,  where 
$\mathcal{A}$  and $\beta$ are the two adjustable parameters of the fitting. The choice of the fitting region is determined by requiring that the fit curve  goes through the data for $L=600$ as well.   This procedure yields $\beta =0.25 \pm 0.02$ for
the   critical exponent that governs the vanishing of the order parameter $\rho$ at the critical point. The resulting fit curve is also shown in Fig.\ \ref{fig:2} using a linear scale.

\begin{figure}[!ht]
  \begin{center}
\includegraphics[width=0.48\textwidth]{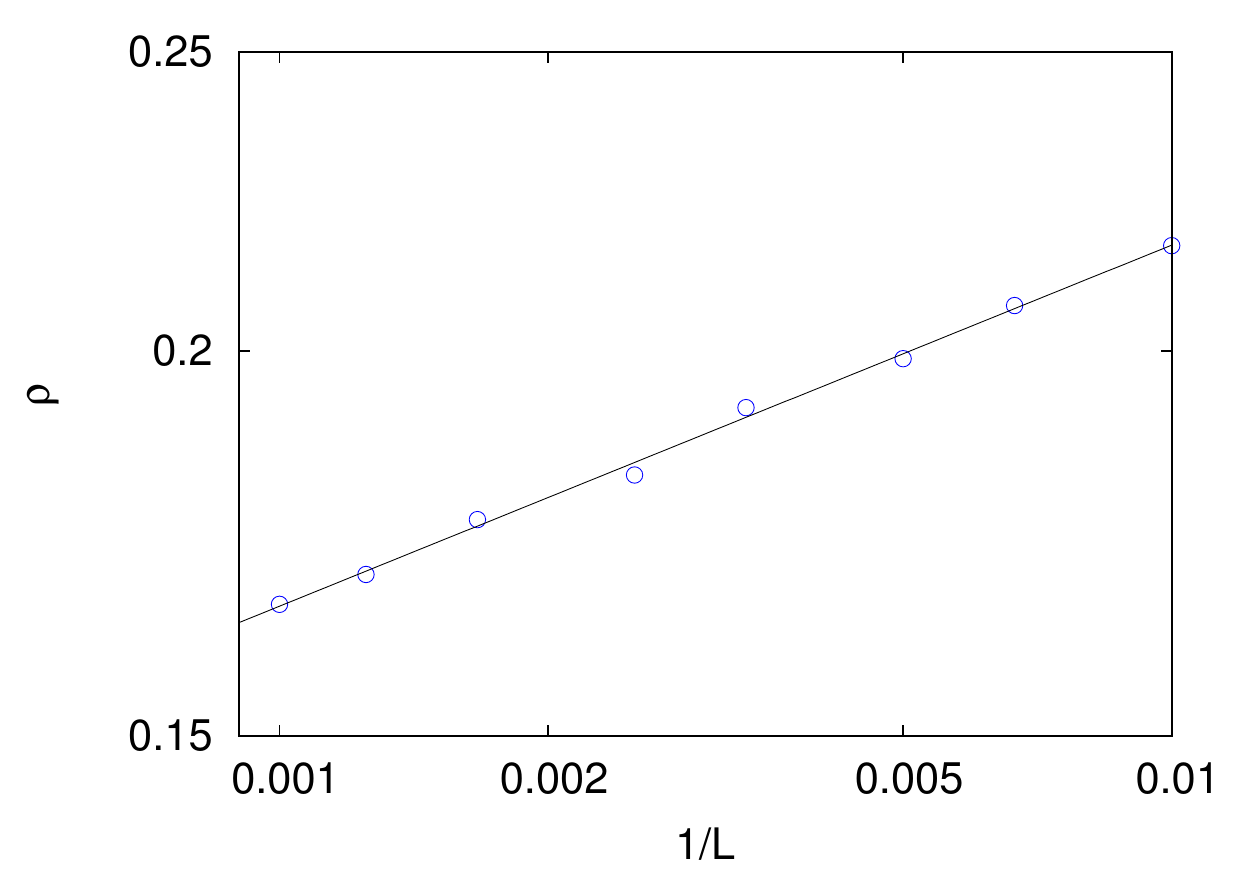}
  \end{center}
\caption{Log-log plot of $\rho$ against the reciprocal of the linear lattice size at the critical point $q_c^0  \approx 3.1$. 
 The curve fitting the data  is  $\rho = \mathcal{B} L^{\beta/\nu^0}$ with $\mathcal{B}= 0.371 \pm 0.005$ and $\beta/\nu^0= 0.117 \pm 0.002$.
The error bars are smaller than the symbol sizes.  }
\label{fig:4}
\end{figure}
 
\begin{figure}[!ht]
  \begin{center}
\includegraphics[width=0.48\textwidth]{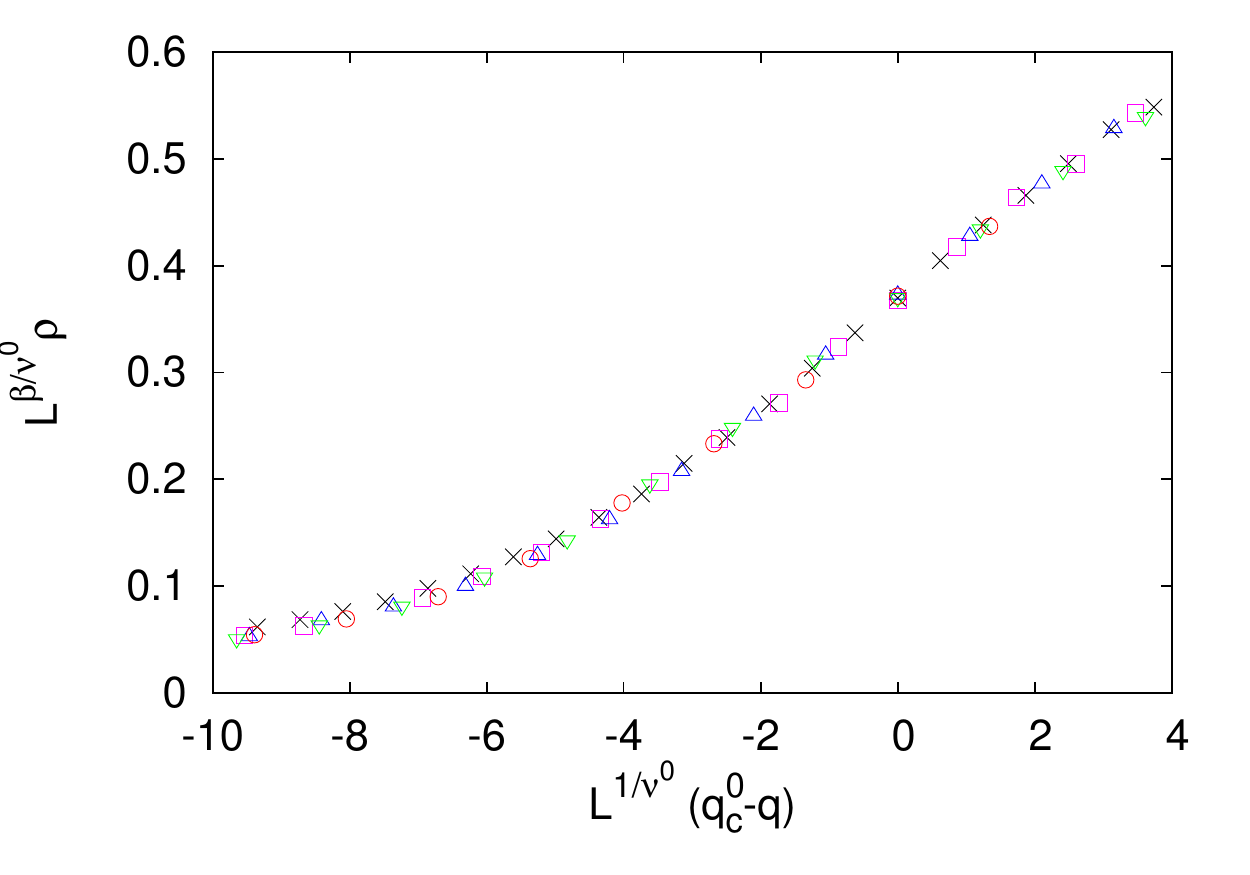}
  \end{center}
\caption{ Scaled order parameter  against the scaled distance to the critical point  for the square lattice ($p=0$) with
linear  size $L =  200 $ ($\times$), $L=400$ ($\square$), $L=600$ ($\triangle$), $L=800$  ($\triangledown$), and $L=1000$ ($\bigcirc$). 
The error bars are smaller than the symbol sizes.
 The parameters are $q_c^0=3.10$, $\beta/\nu^0 =0.117$ and $\nu^0 = 2.1$. }
\label{fig:5}
\end{figure}

 According to the scaling relation  (\ref{scal1}), $\rho$  must decrease to zero as the power law $\rho \sim L^{-\beta/{\nu^0}}$ at $q= q_c^0$ and we explore this fact in Fig. \ref{fig:4}, where $\rho$ is plotted against $1/L$ in a log-log scale, to determine the  ratio 
$\beta/\nu^0 = 0.117 \pm 0.002$. Finally we are now  in position to estimate the exponent $\nu^0= 2.1 \pm 0.1$. The goodness of
our estimates of the critical exponents as well as the judiciousness of   borrowing the location  of the  critical point from \cite{Peres_15}  can be assessed   by checking whether the  scaled quantity $ L^{\beta/{\nu^0}} \rho$ is independent of the lattice size $L$ when plotted against the  scaled distance to the critical point $L^{1/{\nu^0}} \left ( q_c^0  -  q \right )$ as predicted by eq.\ (\ref{scal1}). This is shown in Fig.\ \ref{fig:5}.
The quality of the collapse of the data for distinct lattice sizes  supports heartily our estimates of the critical exponents.

 The two critical exponents $\beta \approx 0.25$ and $\nu^0 \approx 2.1$ that determine the behavior of the order parameter $\rho$ in the critical region of  Axelrod's model  set its continuous nonequilibrium phase transition  apart from the known universality classes  of  nonequilibrium  lattice models \cite{Marro_99}. This is probably due to the static nature of the transition and to the existence of infinitely many  absorbing configurations in both --  ordered and disordered -- phases.
 
 \subsection{Random networks}\label{sec:rand}
 
\begin{figure}[!ht]
  \begin{center}
\includegraphics[width=0.48\textwidth]{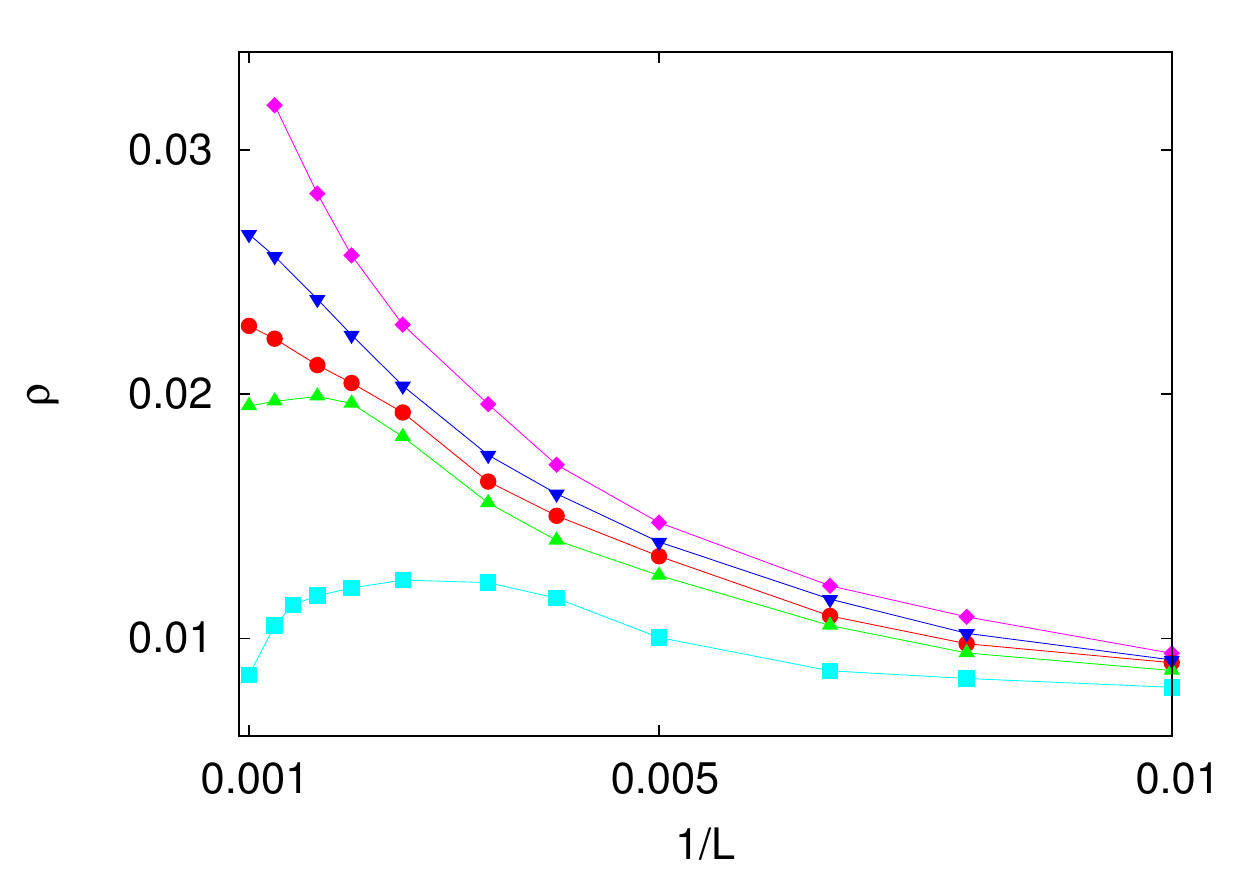}
  \end{center}
\caption{(Color online) Order parameter $\rho$ against the reciprocal of $L$ for random networks and 
 $q =27.0$ ($\blacklozenge$), $27.1$ ($\blacktriangledown$), $27.15$  ($\CIRCLE$), $27.2$ ($\blacktriangle$), and $27.5$  ($\blacksquare$).
 The critical point is $q_c^1 = 27.175 \pm 0.125$. 
The error bars are smaller than the symbol sizes and the lines are guides to the eye. }
\label{fig:6}
\end{figure}

To study the phase transition for random networks generated by the limit of sure rewiring (i.e., $p=1$) of the Watts and Strogatz algorithm \cite{Watts_98} we need first to  obtain a good estimate of the location of the critical point $q_c^1$. According to Fig.\ \ref{fig:1}, in the limit $L \to \infty$ one expects that     $\rho \to 0$ for $q > q_c^1$ and  $\rho \to \rho \left ( q \right ) > 0$ for $q <  q_c^1$, so that the order parameter jumps from $0$ to 
 $\rho_c^1$  at   $q = q_c^1$.  Hence in order to determine $q_c^1$, in Fig.\ \ref{fig:6}  we plot $\rho$ against $1/L$  for several values of $q$ in the critical region. A rough estimate of this region is provided by Fig.\ \ref{fig:1}, which also offers  a good estimate
 for the jump at the critical point, $\rho_c^1 \approx 0.18 $. At $q = q_c^1$ the order parameter is independent of $L$ (hence the intersection of the data for different $L$ at the critical point), i.e.,  $\rho$  exhibits a plateau as  $1/L \to 0$.  Although  such plateaus exist for all $  q < q_c^1$ since $\rho \to \rho \left ( q \right ) $ for large $L$, there is no risk of confusing the two types of plateaus because the range of $\rho$ shown in Fig.\ \ref{fig:6} is about one order of magnitude smaller than $\rho_c^1$. Visual inspection of Fig.\ \ref{fig:6} indicates that $27.15 < q_c^1 < 27.2$ so we estimate $q_c^1 = 27.175 \pm 0.125$.
 
Figure \ref{fig:6} also illustrates the strong finite size effects that hinder the analysis of the order parameter $\rho$ in the critical region. Consider, for instance,  the data for $q=27.5$. If our analysis were restricted to $L \leq 500$ that value of $q$  would be a good candidate for the critical point, because of the plateau observed in the range $200 < L < 500$.  In fact, the data for $p=1$ shown in Fig.\ \ref{fig:1} cross  at about $q=27.5$. Hence to probe the correct critical behavior of $\rho$ we must consider random networks with $L > 600$.  This contrasts to our findings for the square lattice, for which the analysis of rather small networks (e.g., $L=200$) yields  useful information about the critical behavior (see Figs.\ \ref{fig:3} and \ref{fig:5}).

We focus now on the characterization of  the sharpness of the threshold, i.e., the range of $q$ about $q_c^1$ where the threshold  features persist. To achieve that we will assume that the critical region shrinks to zero like $L^{-1/\nu^1}$ as $L \to \infty$, i.e., we will assume that the order parameter is described by the expression $\rho = f_1 \left [ L^{1/{\nu^1}} \left ( q_c^1 -  q \right ) \right ]$ in the critical region. Here $f_1$ is a continuous function   such that $ f_1 \left ( x \right ) \to 0 $ for $x \to - \infty$   and $ f_1 \left ( x \right ) \to \rho_c^1 $ for $x \to \infty$.  
Our approach is in the same spirit of the finite-size scaling of combinatorial problems \cite{Kirkpatrick_94,Monasson_99} (see  \cite{Campos_98} for a similar study in the context of the quasispecies model), for which there is no geometric criterion for defining a quantity analogous to a correlation length, and so the success of the method in accounting for the size dependence of the order parameter $\rho$ cannot be attributed to the divergence of  a correlation length and the consequent onset of a second order phase transition.  In addition, we recall that even  the parameter $L$ has no geometric interpretation for  random networks.

\begin{figure}[!ht]
  \begin{center}
\includegraphics[width=0.48\textwidth]{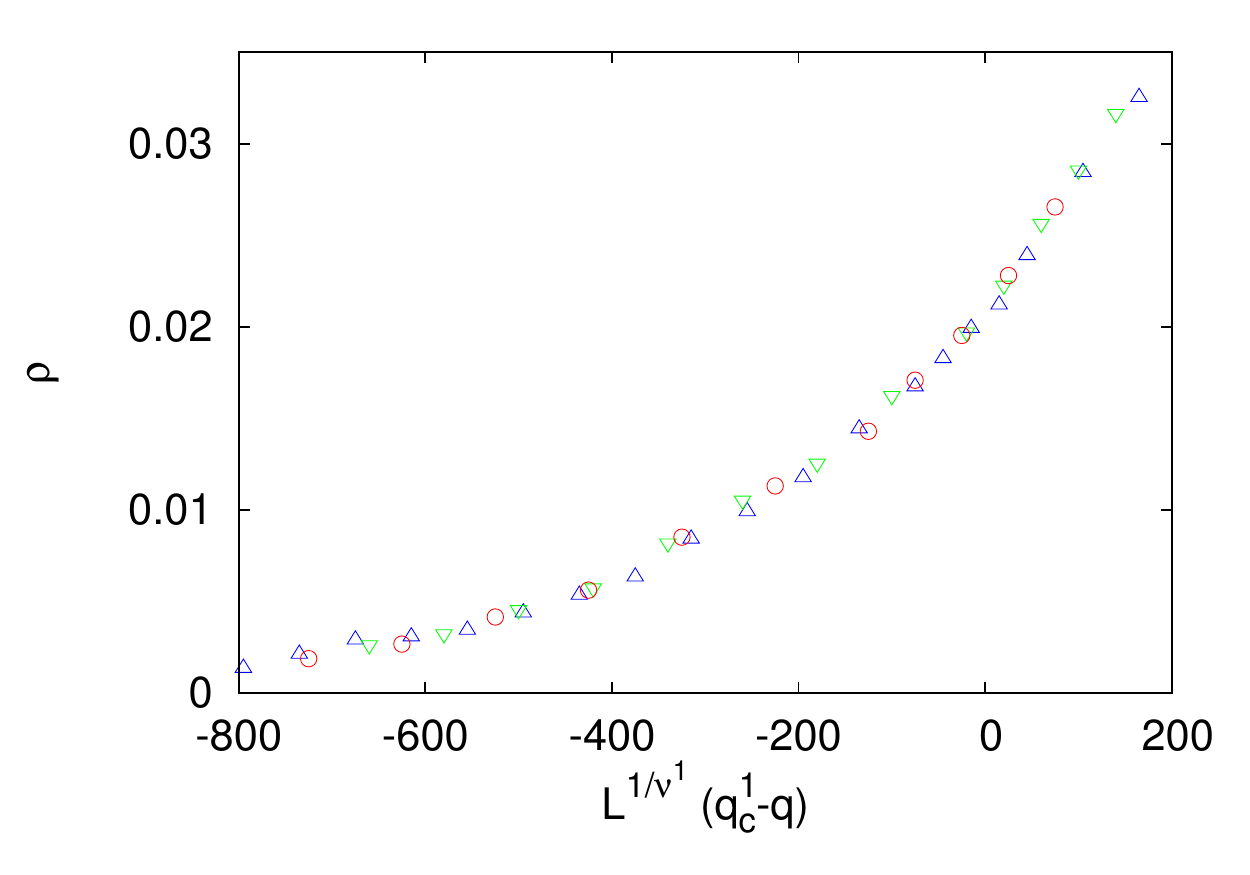}
  \end{center}
\caption{ Order parameter  against the scaled distance to the critical point  for random  networks ($p=1$) with
$L=600$ ($\triangle$), $L=800$  ($\triangledown$), and $L=1000$ ($\bigcirc$). 
The error bars are smaller than the symbol sizes.
 The parameters are $q_c^1=27.175$,  and $\nu^1 = 1.0$. }
\label{fig:7}
\end{figure}
 
 Figure \ref{fig:7}   confirms that, in the critical region, the order parameter $\rho$ is a   smooth function of the properly rescaled distance to the critical point.  The estimate $\nu^1 = 1.0$ for the critical exponent was obtained by requiring that the data for distinct $L$ collapse into a single curve, the function $f_1  \left ( x \right )$. In particular,  we find $f_1 \left ( 0 \right ) \approx 0.021 \ll \rho_c^1$, as expected  (see Fig.\ \ref{fig:6}).  The uncertainty of the estimate of  $\nu^1$   can be  evaluated using the same procedure, i.e., by gauging  the quality of the data collapse as the exponent departs from $\nu^1 = 1$. We find very poor data collapses for exponents outside the range  $\nu^1 = 1.0 \pm 0.05$
 (data not shown). This figure reveals also our difficulty to obtain reliable results in the ordered phase $q < q_c^1$ due to the very long convergence times \cite{Biral_15}. We conclude then that the sharpness of the transition increases with  the square root of the number of agents, $N^{1/2}$.

 \subsection{Small-world networks}
 
 We turn now to the analysis of the discontinuous transition for small-world networks with rewiring probability $p=0.1$.  Our aim is to verify whether the exponent $\nu^p$, and hence the sharpness of the threshold,  is influenced by the value of $p$.  Figure \ref{fig:1} indicates that in the limit $L \to \infty$  the order parameter jumps from $\rho_c^{0.1} \approx 0.22 $ to $0$ at the threshold. The procedure used to determine the critical point is the same as that illustrated in Fig.\ \ref{fig:6} for random networks and yields $q_c^{0.1} = 14.40 \pm 0.05$. The finite size effects are also very strong in this case and so it is also necessary to resort to large networks, say  $L > 500$, to assess the critical region.
 
\begin{figure}[!ht]
  \begin{center}
\includegraphics[width=0.48\textwidth]{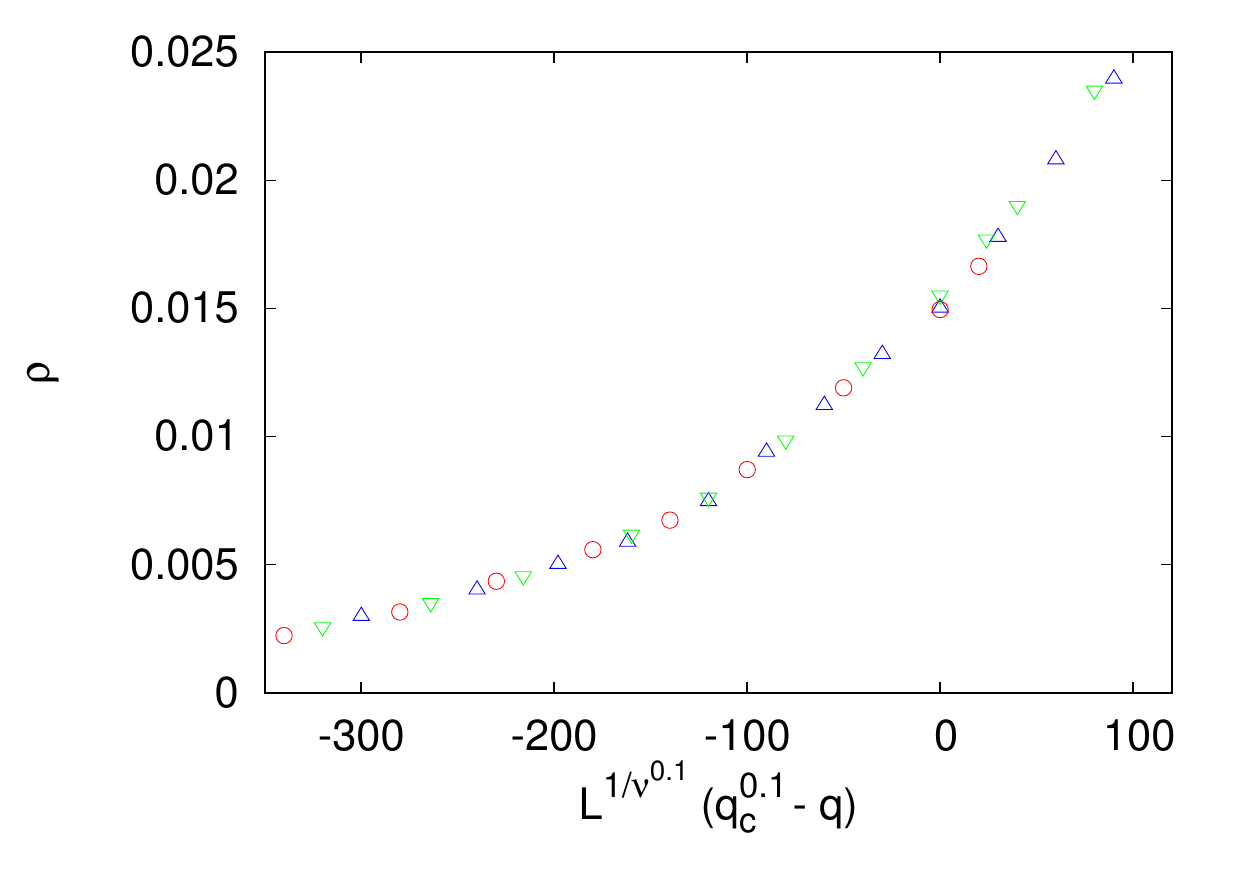}
  \end{center}
\caption{ Order parameter  against the scaled distance to the critical point  for small-world networks with $p=0.1$ and
$L=600$ ($\triangle$), $L=800$  ($\triangledown$), and $L=1000$ ($\bigcirc$). 
The error bars are smaller than the symbol sizes.
 The parameters are $q_c^{0.1}=14.40$,  and $\nu^{0.1} = 1.0$. }
\label{fig:8}
\end{figure}

 In Fig.\ \ref{fig:8} we examine the assumption that the order parameter satisfies $\rho = f_{0.1} \left [ L^{1/{\nu^{0.1}}} \left ( q_c^{0.1} -  q \right ) \right ]$  in the critical region. As before, $f_{0.1}$ is such that $ f_{0.1} \left ( x \right ) \to 0 $ for $x \to - \infty$   and $ f_{0.1} \left ( x \right ) \to \rho_c^{0.1} $ for $x \to \infty$. The best collapse of the data for distinct $L$   is obtained with $\nu^{0.1} = 1.0 \pm 0.05$ and it is shown in the figure.   The uncertainty of the exponent is estimated as before, i.e., by varying $\nu^{0.1}$ around  $1.0$ and gauging the quality of the resulting data collapse. In addition, Fig.\  \ref{fig:8} shows that  $f_{0.1} \left ( 0 \right ) \approx 0.015$, implying that  the data for different  $L$ intersect at $\left (14.40, 0.015 \right)$ in the plane $\left ( q,\rho \right)$.
 
We have also considered values of rewiring probability  down to $p=10^{-4}$ and verified that  the order parameter $\rho$  when  plotted against $q$ for  different values of $L$ (see Fig.\ \ref{fig:1}) always cross at some value $q_c^p$, thus signalling the   existence of
a discontinuous transition between the ordered ($\rho \geq \rho_c^p$) and the disordered ($\rho =0$) phases. Although we have   estimated the sharpness of the threshold for $p=1$ and $p=0.1$ only,  the  observed irresponsiveness of the exponent $\nu^p$ to this tenfold decrease of $p$  prompt us to conjecture that this exponent is  a universal feature  of the discontinuous transitions of Axelrod's model with two cultural features ($F=2$) in  small-world networks.


 \section{Conclusion}\label{sec:conc}
 
Axelrod's model exhibits a good balance between simplicity and realism, being considered a choice model  to   study collective social phenomena  from a quantitative perspective \cite{Goldstone_05}. For instance,  Axelrod's model has proven useful to study the effects of global media on the polarization of public opinion  \cite{Shibanai _01,Avella_10,Peres_11,Reia_16}  as well as to understand  the emergence of a collective intelligence from the local interactions between dumb agents \cite{Kennedy_98,Fontanari_14}. However, rather than explore the riveting applications of Axelrod's model
in social and political science \cite{Axelrod_97}, here we focused on the not less enthralling contributions of Axelrod's model to statistical physics \cite{Castellano_09}.

In particular, our  aim in this contribution was to  offer a quantitative characterization of the critical behavior of the order parameter $\rho$, which measures the fraction of agents, sites or nodes that belong to the largest cultural domain of the absorbing configurations. Such quantitative analysis is feasible only for the Poisson variant of Axelrod's  model \cite{Castellano_00}, since we need to probe the very close vicinity of the critical point in order to compute the critical exponents that characterize the order parameter.
Such detailed study is lacking even for the familiar square lattice \cite{Castellano_00,Peres_15}, as pointed out in Section \ref{sec:Intro}. 

Here we considered a family of small-world networks for which the square lattice and the random networks are extreme limits  of the rewiring probability $p$. The larger the value of $p$, the greater the number of long-range links in the network  \cite{Watts_98}. We found that for $p>0$ the order parameter $\rho$ exhibits a discontinuity at the critical point $q_c^p$ and that $q_c^p$ increases with increasing $p$, indicating that the long-range links enlarge the domain of the ordered phase. 
More pointedly, we found that the transition is discontinuous down to $p=10^{-4}$ and so we conjecture that it is discontinuous for all $p>0$. 
This is similar  to the findings for  the Ising  model in small-world ring lattices, which show  that a ferromagnetic ordered phase exists for any finite value of $p$ \cite{Barrat_00}. We note that these qualitative points were already made in Ref.\ \cite{Klemm_03a}, whose authors studied Axelrod's model in several complex networks, small-world networks included, for integer, uniformly distributed  initial cultural states  and $F=10$. 

Interestingly,   in his  original paper  Axelrod speculated  that the introduction of random long-range interactions would eliminate the  culturally fragmented  absorbing configurations \cite{Axelrod_97}.  Our analysis  shows, however, that even in the extreme case of  random networks   those  disordered configurations persist, provided the disorder of the initial configurations is sufficiently high, i.e.,
$q > q_c^1 \approx 27.17$ (see also  \cite{Klemm_03a}). Nonetheless, in agreement with Axelrod's hunch,  long-range links  do favor  ordered configurations as indicated by the increase of $q_c^p$ with the rewiring probability $p$. This is so because the rewired links provide short-cuts to connect distant sites on the square lattice and, as a result, the walls (i.e., links connecting agents that do not interact because of their antagonistic cultures) can easily be 
circumvented by the dominant culture. In fact, this is also  the reason the ordered phase  is absent in the  Poisson variant of the one-dimensional Axelrod model for $F=2$, whereas  it is present in the  two-dimensional model considered here  \cite{Vilone_02}. The switch of 
the transition from continuous to discontinuous when long-range links are introduced has to do with the suppression of fluctuations
by those links, which makes the model more mean-field-like than the short-range  model  (see  \cite{Herrero_02,Sanchez_02} for the  discussion of  similar findings in the context of the Ising and voter models).

The distinctive aspect of our work on small-world networks is that we were able to characterize the critical region, i.e., to estimate the sharpness of the threshold through the critical  exponent $\nu^p$ (see Section \ref{sec:rand}). Our finding that $\nu^p \approx 1$ for $p=1$ and $p=0.1$  implies that
the sharpness of the discontinuous transition increases with $N^{1/2}$ where $N$ is the number of agents. We expect that this result holds true for all $p > 0$. In fact,  if the value of $\nu^p$ were to depend on $p$, this dependence should be observed with the  tenfold decrease  of $p$
shown in Figs.\ \ref{fig:7} and \ref{fig:8}. 

Our findings regarding the critical behavior of $\rho$ in the square lattice for which the nonequilibrium phase transition is continuous \cite{Castellano_00,Peres_15} corroborate  its unique character: the critical exponents $\beta = 0.25 \pm 0.02 $ and $\nu^0 = 2.1 \pm 0.1$  set it apart from the known universality classes  of  nonequilibrium  lattice models \cite{Marro_99}. This may be due to the  distinctive 
static character  of the transition, which  separates two types of absorbing configurations that differ on their distributions of domain sizes.

The manner we produced the  discontinuous transitions in Axelrod's model  with $F=2$  cultural traits was through the introduction of long-range links on the basal square lattice. However, there is another, perhaps more natural, way to produce those transitions in the square lattice, namely, by increasing the number of cultural traits beyond $F=2$ \cite{Castellano_00}. Although this approach would  result in a considerable raise on the  computational cost to simulate Axelrod's model, the unveiling of the  dependence of the exponent $\nu$ on $F>2$ might be worth the cost since such study would conclude the full characterization of the nonequilibrium phase transition of Axelrod's  model in the square lattice.

\bigskip

\acknowledgments

The research of JFF was  supported in part by grant
15/21689-2, S\~ao Paulo Research Foundation
(FAPESP) and by grant 303979/2013-5, Conselho Nacional de Desenvolvimento 
Cient\'{\i}\-fi\-co e Tecnol\'ogico (CNPq).  SMR  was supported by grant  	15/17277-0, S\~ao Paulo Research Foundation (FAPESP).
This research used resources of the LCCA - Laboratory of Advanced Scientific Computation of the University of S\~ao Paulo.

\end{document}